# On a sub-quantum mechanical interpretation of the Dirac theory

## M. R. Mahdavi

629 Shariati Ave., Tehran, Iran

e-mail: Mahdavi@hotmail.com

In this paper we will attempt to show that the Dirac theory lends itself to an interpretation in terms of a unified sub-quantum mechanical field theory where, the fundamental force fields are weak electric and weak magnetic fields. We present an electron model based on some of the predictions of the Dirac theory and justify our arguments by showing that the model not only yields the correct value for the g-factor but also gives us the very small correction to the magnetic moment as obtained in the renormalization programme of quantum electrodynamics. The model also allows us to give a new meaning to the negative energy state of an electron. The general outlines of the unified field theory are given and the connection between fluctuations at the Planck and Compton scales is presented in terms of concrete physical events.

## 1. Introduction

    The development of a true sub-quantum mechanical theory, namely, a theory that goes beyond a mere interpretation of quantum mechanics, requires three stages of development. First, one has to form in ones mind an overall picture, albeit a vague one, of sub-quantum mechanical physics. Second, one has to discover the nature of the hidden parameters; partly as a result of the exigencies of the problems that one is faced with, and partly from inferences drawn from theory and observational results. Third, having found the physical make-up of the sub-quantum particles, one should then develop the fundamentals of sub-quantum mechanics and present a more detailed and acceptable picture. In particular one should resolve the problems that one came across in the first and second parts of this programme, but due to a lack of knowledge had to conveniently side step.

    The aim of the present paper is to carry out the first and to some extent the second part of the above programme. In a previous paper [1] we demonstrated that the unspecified hidden

physical property, advocated by the proponents of local realistic models [2-4], amounts to a non-spherical electric charge distribution for the electron. The aims of that paper were; (1) to demonstrate that the concept of "efficiency loophole" cannot be advocated in support of a local realistic interpretation of quantum mechanics, since a non-spherical charge distribution automatically requires alongside it superluminal signalling, and, (2) to briefly consider the efficacy of non-spherical charge distribution in a non-local realistic interpretation of quantum mechanics. There we concluded that a non-spherical charge distribution might be the root-cause of quantum mechanics and that, it is an idea that deserves to be further investigated.

In the present paper we wish to advance the idea that the Dirac theory [5], not only predicts a non-spherical charge distribution for the electron, but that if one takes the view that all its predictions are essentially correct, then one would be forced to move into a sub-quantum mechanical theory. We justify our arguments by deriving the correct value for the magnetic moment of the electron, and more importantly by getting the correct value of the second order correction to it.

## 2. Review of the Dirac theory and background to the present paper

The followings are most of the immediate predictions that appear in the Dirac theory of the electron:

I. The eigenvalues of a component of the velocity operator are equal to ±1.
II. The components of velocity operator do not commute with one another.
III. A component of velocity is composed of two parts, a constant part corresponding to velocity as it appears in the classical momentum relation, and an oscillatory part whose frequency of oscillation is equal to $\nu_r = 2mc^2/h$.
IV. A position coordinate is also composed of a constant and an oscillatory part. The oscillatory part is imaginary, and the amplitude of oscillations is of the order of $\hbar/2mc$.
V. There are two wave functions for each of the energy states of the electron.
VI. The state vector is double-valued in ordinary 3-space and returns to itself after a rotation through $4\pi$.
VII. In an external magnetic field the electron behaves as if it had a magnetic moment of magnitude $\mu_e = q_e\hbar/2m$. In contrast to the classical situation the g-factor has the abnormal value of 2.

VIII. In an external electric field the electron behaves as if it had an imaginary electric moment $p_{im} = iq_e\hbar/2mc$.

Of the above, prediction I stands in apparent conflict with the special theory of relativity, since it implies that a measurement of the instantaneous velocity of the electron is certain to lead the results ±c. However, by invoking the uncertainty principle, Dirac showed that this conflict might be resolved.

Predictions III and IV seem natural in a quantum mechanical theory, since they associate oscillations with the motion of the electron, as was originally postulated by de Broglie. However, it seems odd that the frequency of oscillations is twice as large as the one proposed by de Broglie. Also the position coordinate is non-Hermitian. It has been argued [6] that this is symptomatic of a non-commutative geometry.

In terms of classical concepts predictions VI and VII are unnatural, but must be accepted as being correct, since they are in keeping with the observational results.

Predictions II and V have been attributed to an inadequacy of the Dirac theory [7] in representing a true physical picture of the electron since; II means that if one measures a component of velocity in a given direction one cannot at the same time obtain any information about the components of velocity in other directions, while V amounts to a redundancy in the number of the wave functions necessary to describe each energy state of the electron.

Finally, VIII has been dismissed altogether on the grounds that an imaginary electric moment cannot correspond to a real physical entity.

We now adopt the view that the Dirac theory is a very accurate theory, representative of an overall average of a number of phenomena that collectively manifest themselves as the particle, the electron. In the present paper, except for V and VIII, we shall make use of these predictions and shall present an electron theory based on them. In a subsequent paper we shall show that V and VIII are also correct predictions.

## 3. An electron model based on the Dirac theory

We propose the following model of the electron as a way of explaining the non-commuting character of the components of the velocity operator. We assume that the Coulomb charge of the electron is in the form of an annular band of radius d and thickness t, with β being the angle subtended by t at the centre of the annular band (**Fig. 1**).

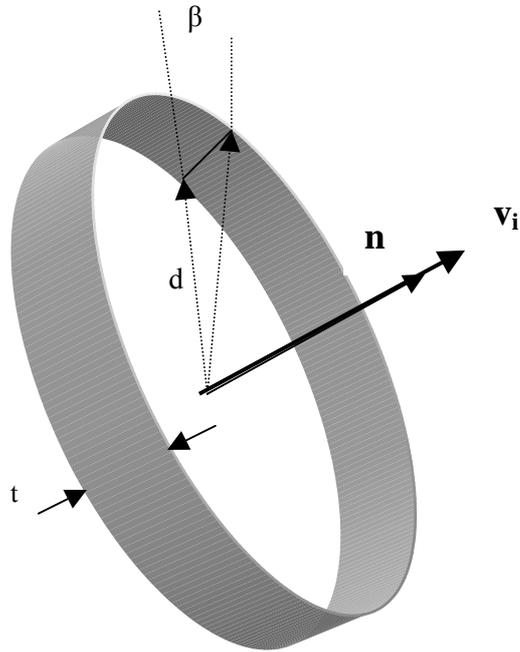

**Fig. 1**

If as shown in **Fig. 1**, the electron were to be set into motion, in such a way that $\mathbf{v}_i$ the velocity vector coincided with **n**, a normal vector to the plane bisecting the band into two equal annular bands, then the electrostatic field of the electron will come into ones view for a short period of time, only when the direction of ones view is almost transverse to the direction of $\mathbf{v}_i$.

Next, consider that the electron has an intrinsic angular momentum of magnitude $\hbar/2$. Since in the Dirac theory a position coordinate has a constant and an oscillatory part we argue that, in the case of the electron at rest, the electron undergoes some sort of an oscillatory motion along a closed trajectory, for example a circle, and attribute the intrinsic angular momentum to such a motion. Thus, tentatively, we let the annular shaped charge to move in circular motion, with $\mathbf{v}_i$ parallel to **n** as in **Fig.1**.

Now, if one were to observe the electron from any point in space, save from points on the polar axis, one would observe the electrostatic field of the electron for short periods of time when the vector velocity is almost perpendicular to ones direction of view. This model, however, will not yield an isotropic field for the electron, since if the point of observation is on the polar axis then one would be enabled to see the electrostatic field at all times. We overcome this problem by means of prediction VI. Thus as before we let the electron move in circular orbit, but in addition allow **j** the angular momentum vector or to precess about the z-axis at an angle of inclination θ ≥ β at a recessional frequency equal to $\tfrac{1}{2}\omega_r$ where, $\omega_r$ is the rotational frequency. Thus in **Fig. 2**, if at an instant of time the electron is at the position **A** in space then, after a rotation through 2π it will move to the new position **B**. It will now take another full cycle of rotational motion before it returns back to its original position in space. The combined rotational and precessional motions at the indicated frequencies ensure that the frequencies and the durations of observations would be the same from all points of observation in space.

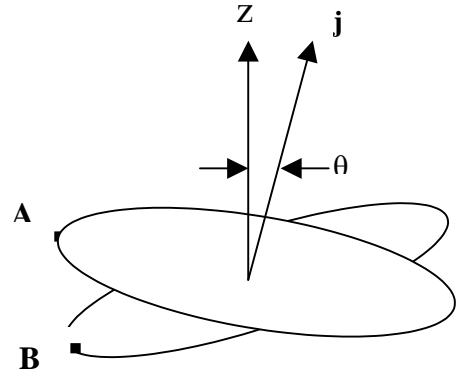

**Fig. 2**

Thus on the average the electrostatic field of the electron attains an isotropic form.

As explained in [1], under normal circumstances the electrostatic field of the electron will now be in the form of superluminal pulses of time duration Δt, arriving at the point of observation with the $C = (1+\beta^{-1}) \cdot c$ and at intervals of $\Delta t' = (1+\beta^{-1}) \cdot \Delta t$. Thus

$$\alpha = \Delta t/\Delta t' = 1/(1+\beta^{-1}) \qquad (1)$$

is the ratio of the period time durations when the electron comes into ones view or disappears from it. In [1] we also argued that α might be equal to the fine-structure constant

$$\alpha = (q_e^2)/4\pi\varepsilon_0 c\hbar \qquad (2)$$

We now adopt the view that this is indeed the case and, with a view to future developments assume that the electronic charge is distributed on an annular band of radius d and thickness t with t/d = α (ignoring the small difference we have put α = β). Therefore, the charge density on the band is

$$\sigma^- = q_e/2\pi\alpha d^2 \qquad (3)$$

Thus, viewed from the directions along which one can see the electron's charge one would see an electrostatic field as if emanating from a charge

$$-Q = -2\alpha^{-1} q_e \qquad (4)$$

In the course of coming discussions it will soon become apparent that an electron is the manifestation of a number of phenomena. Therefore, we should make a distinction between the electron and its particle part. For this reason we shall refer to the particle part of the electron as the electrino.

**4. The instantaneous velocity of the electrino**

In order that we may be consistent in our interpretation of the Dirac theory we have no choice but to accept that, prediction I in fact means that the instantaneous velocity of the electrino, and not the electron itself, is equal to $\pm c$. If this view is correct we must then show why it is that the motion of the electrino with the velocity of light is not in conflict with the special theory of relativity. Two problems that require immediate attention are the followings.

First, if a particle of a finite inertial mass were to move with the velocity of light, then according to the special theory of relativity its mass would attain an infinite value. From this we conclude that the electrino in essence is a zero mass particle, *i. e.*, a tachyon. Next consider that the tachyon cannot of its own volition, as depicted in **Fig. 1**, move in a circular orbit. Thus, a force field $\psi^*$ of electron's own making must act on the tachyon and constrain it to move in circular motion. Therefore, when the electron is subjected to an external force, the $\psi^*$-field will oppose the motion of the tachyon. In this way the initially zero-mass particle shows resistance to motion and as a result has acquired the attribute of inertial mass, which we will equate to m the observed mass of the electron.

Second, a tachyon cannot carry ordinary electric or magnetic charges. For, if it were to have a charge then its electrostatic field in directions transverse to the direction of its motion would attain infinite value. Therefore, any charges initially carried by a tachyon must be of the weak type, *i. e.*, the field lines of a charge on the tachyon are confined to its immediate vicinity and as it were are in the form of permanent attachments to it. We are now forced to the conclusion that the Coulomb charge is not an intrinsic physical attribute of the electron, but that it is created when the tachyon is captured into the electronic motion. That Coulomb charge is not an intrinsic physical attribute of an elementary particle is already indicated in quantum electrodynamics where a creation operator, for example, creates an electron together with its Coulomb field and destroys a positron together with its Coulomb field.

An additional problem now comes to ones mind. What laws govern the motion of the electrino? We will assume that the motion of the electrino is in accordance with the laws of classical physics. In particular we note that in classical physics the motion of a charge gives rise to an induced magnetic field. Here, we are at the stage that the charge is just created. However, we shall assume that there is a magnetic field associated with the motion of the electrino, exactly as given by the classical relation for induced magnetic fields except that, the Lorentz transformations do not apply to the motion of the electrino or its electric or magnetic fields.

## 5. Discontinuous motion

Although we have postulated that a magnetic field is associated with the moving electrino, nonetheless, no essential change has taken place from the classical situation and our model of the electron still yields the classically incorrect value for the g-factor. In order that we may get the correct g-factor we need to modify our electron model in two steps. The first step is to make the motion of the electrino discontinuous in the true sense of the word.

Schematically we represent the events in two dimensions in the manner shown in **Fig. 3**. In the first instance and as the result of action of the creation operator, a tachyon is captured into an electronic orbit at point A, and after having been converted into an electrino, moves along a circle of radius $a_e$, again via the agency of the creation operator. We now make the following assumption and will justify it later on in section 7. We assume that an electrino thus created has only a brief lifetime. That is, we assume that the first electrino after a short period of time corresponding to motion with the velocity $c$ along an arc length $\Delta s$ = AB

$$\Delta s = \alpha a_e \qquad (5)$$

once again becomes a free particle and falls back into the vacuum. In keeping with prediction IV we let the radius of circular motion to be equal to

$$a_e = \hbar/2mc \qquad (6)$$

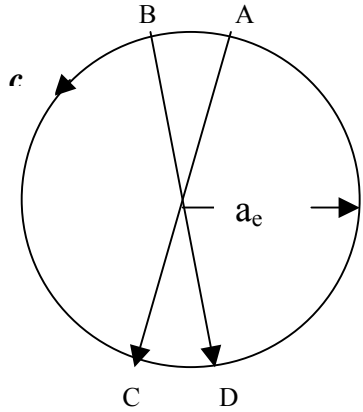

**Fig. 3**

Now the frequency of motion is equal to $\nu = 2mc^2/h$ as in prediction IV, and assuming that the acquired mass of the electrino is equal to $m$, the inertial mass of the electron, then the value of angular momentum attains the correct value of $\hbar/2$.

Next, consider that if the first electrino has a finite lifetime then, at the instant that it reverts back into the vacuum, another electrino must be created at a nearby point in space and forced to move along an arc length as given by eq. 5 This is mandatory in order that we do not violate the law of charge conservation. This situation can come about if a field identical to the original creation operator emanates from the first electrino. Considering the fields associated with an electron it is reasonable to assume the $\psi^*$-field is in fact an electromagnetic one since the gravitational field is too weak for the considered task. Clearly, the $\psi^*$-field must have three components in three mutually perpendicular directions rθz, and these components must act on some hidden physical attributes of the tachyon in order that it can be kept in circular orbit.

Now, since electromagnetic fields travel away from their source along straight lines we may expect that the second electrino will be created at a point C (**Fig. 3)** and subsequently will move over an arc length CD = Δs. However, although the idea is essentially correct, nevertheless, it needs to be somewhat modified, since this model cannot overcome the problem mentioned at the beginning of this section.

## 6. The hidden space warp, the g-factor, and a new spin

If a new electrino is always created at a distance of $2a_e$ from the source of the $\psi^*$-field it then follows that at ($\mathbf{E}_i$, $\mathbf{B}_i$), the components of the electric and magnetic fields in the $\psi^*$-field, must attain certain values ($\mathbf{E}_{i0}$, $\mathbf{B}_{i0}$) at this distance. Now, as before we let the first electrino moves along an arc $\Delta s_a$ of a circle of radius $a_e$. The $\psi^*$-field emanating from this electrino (schematically represented as a ray in the figure) when it was the point A moves inward with velocity $C' = \alpha^{-1}c$, not along a diameter but at a small inclinaton to it (**Fig. 4**). Having travelled a distance ≈ $a_e$ the $\psi^*$-field is close to the centre of the circle. We now assume that at the centre of the circle there is a space warp, resembling a black hole, and that the $\psi^*$-field arrives at a point very near to the black hole horizon. Here, we emphasize that the space warp is not due to the presence of a particle; rather, it is a local alteration in the space-metric that must continually be replenished, otherwise it will simply fade away (see sect.8). Now, a number of the electric and magnetic field lines, for example those closest to the black hole horizon, are drawn towards the black hole. As a result the direction of the travel of the $\psi^*$-field is now altered, and having looped around the black hole it starts to move radially outwards and after having travelled a further distance equal to $a_e$ it creates a new electrino at point B. This sequence of events is repeated for all $\psi^*$-fields emitted during each point of motion of the first electrino. The net result is that the arc length over which the next electrino moves is projected not onto a diametrically opposite position, but rather, onto a position near the previous arc length. Thus the arc lengths over which successive electrinos move correspond to a single arc length moving with a velocity $C'$ over a circle of radius $a_e$.

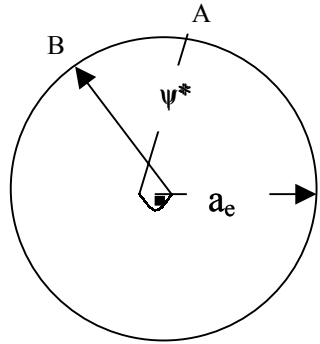

**Fig. 4**

Now, in order that the measured charge and the magnetic moment of the electron may have their correct values $-q_e$ and $q_e\hbar/2m$, respectively, we must put the value of $C'$ at

$$C' = \pi(1+ \alpha/2\pi)c \qquad (7)$$

The choice of the value for $C'$ was made from the following considerations. If we put $C'$ equal to $\pi c$ then, in appearance an electrino and also an arc length move with a velocity $2c$ along a diameter to the diametrically opposite position. In the case of the electron at rest, this means that if we observe (or do not observe) an electrino moving over a full arc length then we are certain to observe (or not to observe) it moving over a full arc at the diametrically opposite position. For this reason we must choose a value for $C'$ such that; for all observers, the sum of the observable arc lengths $\Delta s_a$ and $\Delta s_b$, at two diametrically opposite positions, would always add up to $\alpha a_e$, for example, $\Delta s_a = (1-\varepsilon)\alpha a_e$ and $\Delta s_b = \varepsilon\alpha a_e$ where $\varepsilon$ takes on any values within the range 0 to 1. The additional velocity $\Delta c = (\alpha/2\pi)c$ in eq. 7 were put in for this purpose and this value of the $C'$ ensures that the ratio $\Delta t/\Delta t'$ will always have the correct value $\alpha/2$ for all observers

The net result of all this for the case $\varepsilon \ll \alpha$, is depicted in **Fig. 5**. The electrino appears and moves over the arc length $\Delta s_a = (1-\varepsilon)\alpha a_e$. In appearance it moves with a velocity $2c$ to the diametrically opposite position where it moves over an arc length $\Delta s_b = \varepsilon\alpha a_e$ and subsequently with the same velocity back to its original position. Thus, in this two-dimensional picture, the appearance time of an electrino is equal to $\Delta t \approx \alpha a_e/c$ while its disappearance time is equal to $\Delta t' \approx 4a_e/2c$. Therefore the ratio ($\Delta t/\Delta t'$) is equal to $\alpha/2$ and since, when the electrino is observed it presents a charge of $-2\alpha^{-1}q_e$, it follows that the average charge of the electron will be equal to $-q_e$. Now the frequency of motion of the charge is $\nu \approx 2c(1+\alpha/2\pi)/4a_e$, the value of the current is $i \approx \nu 2\alpha^{-1}q_e$, and the area enclosed by the electrino current is equal to $A \approx \alpha(a_e)^2$. Therefore the value of the magnetic moment of the "circulating" electrino is equal to

$$\mu_e \approx [2c(1+\alpha/2\pi)/(4a_e)](2\alpha^{-1}q_e)(\alpha a_e^2) \qquad (8)$$

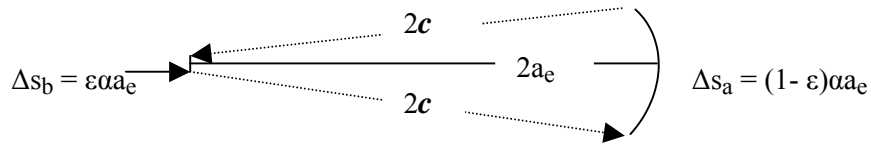

**Fig. 5**

Inserting for the value of $a_e$ from eq. 6 into eq. 8 we obtain a value for the magnetic moment equal to

$$\mu_e \approx (q_e\hbar/2m)[1+(\alpha/2\pi)] \qquad (9)$$

Thus, we not only have got the correct value for the magnetic moment but have also got the correction $\Delta\mu_e = (q_e\hbar/2m)(\alpha/2\pi)$ as also predicted by renormalization theory.

In the three-dimensional case events will proceed more or less in the same way except that, as depicted in **Fig. 1**, the electrino after one cycle of rotation does not return to its original position, *e. g.,* its z-coordinate oscillates between the values $z_0 \pm dz$ with $dz \sim \alpha z_0$.

From the preceding discussion it will be seen that two spins are required for a correct description of the electron. There is a ½-spin associated with the rotational motion of the electrino. In addition there is a spin s', which arises purely as a result of the presence of a local space-time curvature. In the above treatment (**Fig. 4**) we arbitrarily allowed the $\psi^*$-field to approach the black hole horizon from left side. Thus for a spin-up electron s' will have the same direction as the angular momentum vector. On the other hand, had we let the $\psi^*$-field approach the black hole horizon from the right side then in appearance nothing would have changed, but the direction of s' would have been reversed. However, there is nothing here to tell us the s' state of the electron.

**7. The hidden "positron"**

Dirac explained the negative energy state of the electron in terms of his hole- theory. His explanation was seemingly justified by the discovery of the positron. However, we take a different view about the negative energy state.

Consider that we have the $\psi^*$-field that in appearance continually bounces from one electrino to another. In effect it is a packet of electromagnetic field trapped inside the electron and in a later paper we will show that it is in fact an electromagnetic mesic field. However, in so far as the effect that it produces it corresponds to an electrostatic field due to a charge $+q_e$ always at the position where the electrino is situated. Thus, if we were to force an electrino to move into circular motion as a result of the presence of a positive charge $+q_x$ then the following relation must hold

$$(mc^2/a_e) = (2\alpha^{-1} q_e \cdot q_x)/(4\pi\varepsilon_0)(2a_e)^2 \qquad (10)$$

which on simplifying reduces to

$$q_x = q_e \qquad (11)$$

## 7. The outlines of the sub-quantum unified field theory.

Now we may ask, if a tachyon can only carry weak type of electric or magnetic charges then, how its long range fields are created? We believe that Michael Faraday supplied the answer to this question a long time ago. Faraday considered space to be filled by tiny dipoles and envisaged that electric fields are conveyed from one point to another in space by a process of polarization. It was on the basis of Faraday's ideas that Maxwell introduced the concept of displacement current and constructed his electromagnetic theory [8]

How Faraday's idea can be translated into practical terms will be discussed in a separate paper. Here, we wish to consider in general terms a few points that follow from this idea. First, consider the electrino at a point in its circular motion. It has a certain amount of weak electric charge (and possibly weak magnetic one) and the field lines from this weak charge penetrate a limited distance d into space. Thus, plausibly we may expect the weak charge to polarize some of the tiny dipoles within the range of its weak field. At this instance we have a number of filaments of polarized dipoles emanating from the weak charge. Obviously, the tiny dipoles must show resistance to a turning force and after the weak charge has moved away from this region of space they should relax back to their original orientations in space, as otherwise the transmission of signals through space would proceed with infinite velocity. Now we have to assume that the first filament of polarized dipoles, before the dipoles relax back to their initial equilibrium positions, will in turn polarize another filament of dipoles and that in this way and in effect a filament of polarized dipoles will move with a finite velocity through space.

Second, an important point here is that the weak field must fall-off with distance as $1/r^n$, with n itself being an increasing function of r, so that at a distance of $r = d$ the value of the field falls to zero. In other words there is a sharp gradient in the weak field.

Therefore, what the weak field does is, in addition to polarizing the tiny dipoles, to pull them towards it. Clearly, the physical properties of space around the weak charge are now altered. In particular if before the introduction of the weak charge in a given region of space the average distance between the tiny dipoles were equal to $d_0$, it will now be reduced to $d_0(1-\delta)$. Here we have an alteration in the physical property of space, reminiscent of Einstein's theory of gravitation. If this is indeed how gravity is created then it means that the electromagnetic field, wherever it goes in space, brings with it the gravitational field.

Third, we have filled space with dipoles that show resistance to a turning force. Thus, space itself has some sort of inertia. This means that there is a large amount of unobserved mass inside our universe. Note that the fact that space itself has inertia in no way means that, a body moving through space will gradually loses its momentum. For, the motion of a body is determined by the path of the $\psi^*$-fields and in reality there is no motion corresponding to motion as in the classical situation.

Next consider that a $\psi^*$-field always lifts a tachyon out of the vacuum and converts it into an electrino, always at a distance of $2a_e$ from its source. Thus, it is a short-range field whose attractive quality is manifested only at a certain distance. Thus, in one respect at least the $\psi^*$-field resembles the strong field. Should further investigation show that the $\psi^*$-field is in every respect identical to the strong field we then will have a unified field theory where the fundamental force field of nature is the weak electromagnetic field and gravitational, electromagnetic, and the strong fields are merely different manifestations of the weak fields.

## 8. The connection between phenomena at the Planck and Compton scales

Previously the question arose "why should electrinos move over arc lengths all exactly equal to $\alpha a_e$?". In order that we may answer this question let us look at the black hole. It is in reality a space warp that, if left to its own, would fade away in a short time. Thus, it needs to be continually replenished. A gravitational field originating from the electrino itself can only do this replenishment. Now, the only thing that emanates from the electrino and reaches the position of the black hole is the $\psi^*$-field. Therefore, even in the absence of the above outlined the unified field theory, we would have been forced to assume that the $\psi^*$-field, an essentially electromagnetic field, does the job of replenishment. This can only be done if the electromagnetic field has always an associated gravitational field.

Next, consider that the space warp must correspond to a particle having a constant mass. Now, with a fixed arc length $\alpha a_e$ we always have a certain number of filaments of electric field circulating around the centre of the circle. If the arc length is less than $\alpha a_e$ the number of the circulating filaments, and in consequence the mass of the black hole, is reduced and the black hole will start to evaporate. On the other hand if the arc length is increased beyond $\alpha a_e$ then the number of the circulating field line filaments, and in consequence the mass of the black hole, is increased. As a result the filaments, which previously were circulating at the black hole horizon will now be drawn inside the black hole. The field lines will now be trapped inside the black hole up till the time that it starts to evaporate, but we are in no position to state what might happen afterwards. Thus, there is fine balance between what is going on at the Compton scale and the disparately smaller scale of the black hole. These bring us now to the subject of quantum gravity.

In this paper we were forced by necessity to combine gravity and an electron model based on the Dirac theory in order to get the correct value of the g-factor. For some times now Quantum Gravity [9,10] has been the subject of a good deal of theoretical research and there is already some evidence showing that there might be a close connection between quantum mechanics and gravitation [Cf. ref.11 and refs. therein]. Let us now take one particular aspect of a black hole that is of particular relevance to the present discussion. In quantum gravity the minimum scale length is the Planck length

$$l_p = (\hbar G/c^3)^{1/2} \sim 10^{-33} \text{ cm} \qquad (12)$$

Associated with $l_p$ is the Planck mass $m_p \sim 10^{-5}$ gm, which corresponds to a black hole whose Schawrzchild radius is equal to $l_p$. Now, such a black hole would evaporate via the Beckenstein radiation in a time $\sim 10^{-42}$ seconds. This time scale for black hole radiation is more than twenty orders of magnitude smaller than time scales at the Compton scale of elementary particles. It is now evident that at the start we cannot put an electrino in an electronic orbit and hope to replenish the evaporating black hole. Therefore, the connection between the black hole and the electrinos can be maintained, if at the instant that the first electrino appears, there are n filaments of field lines strung together; extending from the position of the first electrino to the position of the black hole, and after looping around the black hole, extend partly in space towards the point that the next electrino is due to appear. Only in this way the fluctuations at the Planck scale can be causally connected to the oscillations at the Compton scale. In regards to this connection at the two disparate scales we draw the attention of the reader to a recent paper by Sidharth [12]. There, Sidharth considers and offers an explanation for the curious fact that in Quantum Gravity and Quantum Super String theories the scale length is the Planck scale, whereas in the physical world the scale length is the Compton scale of the elementary particles.

## 9. Conclusions

   In conclusion we may say that the fact that we were able to derive the correct value for the g-factor and at the same time obtain the small correction to the magnetic moment goes a long way in supporting our contention that the Dirac theory can be explained in terms of a sub-quantum mechanical theory. In regards to the unified field theory, in general terms we made a connection between the four known forces of nature, but clearly it needs to be worked out in more details. Altogether, we may claim that our sub-quantum theory reproduces some aspects of the quantum theory and also removes the conceptual difficulty that one has in regards to the negative energy state of the electron.

**References:**

[1]. Mahdavi, M. R., physics/0211102 (2002).
[2]. Fine, A., *Synthese*, **50** (1982) 279.
[3]. Szabo, L. E., *Found. Phys.*, **30** (2000) 1891.
[4]. Larsson, J.-Å., *Found. Phys.*, **13** (2000) 477.
[5]. P. A. M. Dirac, *The Principles of Quantum Mechanics*, (Oxford Univ. press, 1958) 261.
[6]. Sidharth, B. G., physics/0203079 (2002).
[7]. Schweber, S. S., *Relativistic Quantum Field Theory*, (Harper&Row, 1964) 91.
[8]. Maxwell, J. C., *A Treatise on Electricity and Magnetism*, **vol. II** (Dover, 1954) 431
[9]. Hawking, S., *Commun. Math. Phys.* **43** (1975) 199.
[10]. Horowitz, G. T., gr-qc/0011089 (2000).
[11]. Matone, M., hep-th/0005274 (2002).
[12]. Sidharth, B. G., physics/0208045 (2002)